\documentclass[%
 reprint,
superscriptaddress,
 amsmath,amssymb,
 aps,
prb,
]{revtex4-2}

\usepackage{color}

\usepackage{graphicx, float}
\usepackage{dcolumn}
\usepackage{bm}
\usepackage{comment}
\DeclareMathOperator*{\Motimes}{\text{\raisebox{0.25ex}{\scalebox{0.8}{$\bigotimes$}}}}

\begin{document}

\preprint{APS/123-QED}

\title{Searching for activated transitions in complex magnetic systems}

\author{H. Bocquet}
 \email{h.bocquet@protonmail.ch}
\author{P. M. Derlet}
    \email{peter.derlet@psi.ch}
\affiliation{Condensed Matter Theory Group, Paul Scherrer Institut, CH-5232 Villigen PSI, Switzerland}

\date{\today}

\begin{abstract}
The process of finding activated transitions in localized spin systems with continuous degrees of freedom is developed based on a magnetic variant of the Activation-Relaxation Technique (mART). In addition to the description of the method and the relevant local properties of the magnetic energy landscape, a criterion to efficiently recognize failed attempts and an expression for the step magnitude to control the convergence are proposed irrespective of the physical system under study. The present implementation is validated on two translational symmetric systems with isotropic exchange interactions. Then, in one example, diffusion processes of a skyrmion vacancy and a skyrmion interstitial are revealed for a skyrmion system on a square spin lattice. In another example, the set of activation events about a metastable state of a 2D dipolar spin glass is investigated and the corresponding energy barrier distribution is found. Detailed inspection of the transition states reveals the participation of nearest neighbour pairs affording a simplified analytical understanding.
\end{abstract}

\maketitle


\section{\label{sec:1}Introduction}
Exploring the phase space of a physical system in order to recover statistical information can be achieved either via direct dynamics or from a systematic exploration method that relies on local properties of the energy landscape. Evolving the system under its dynamics may be inefficient when long-time relaxation processes are relevant. In this case, the statistical information is more rapidly obtained by a systematic exploration method. For Hamiltonian mechanics, momentum degrees of freedom are easily computed and factored out from the canonical ensemble averages. This conveniently restricts the exploration to the positional degrees of freedom appearing in the potential energy. Potential energy landscape (PEL) exploration can be realized thermally by sampling microstates with Monte Carlo simulations, or directly by using the local properties of the PEL with saddle point search methods. The latter identify activated transitions between locally stable states whose characteristic timescales can be obtained by other means~\cite{Wigner_1938, Braun_1994, Bessarab2012}.

Technically, saddle point search methods aim at finding transition paths between neighbouring local minimum configurations of the PEL. These transition paths are characterised by the presence of a first-order saddle point, i.e. a stationary point with exactly one unstable direction. An example of this technique is the Nudged Elastic Band (NEB) method, which consists of refining an initial path between two known minima by relaxing in the perpendicular direction to the path~\cite{Wales1990}. Other methods, such as the eigenvector-following method, the Activation-Relaxation Technique (ART) and the dimer method iteratively search for a saddle point and a new neighbouring minimum surrounding a given minimum. These so-called open-ended techniques are relevant to search for an a priori unknown neighbouring local minimum and can even be set to find multiple transition events about an initial configuration. The eigenvector-following method progresses towards saddle points of the PEL by maximizing along one preset eigenvector while relaxing individually along all the others using a modified Newton-Raphson scheme on the force~\cite{Wales1994}. This scheme requires the evaluation of the full eigenspectrum of the PEL Hessian at every iteration, which is numerically expensive. In contrast, ART~\cite{Barkema1996, Barkema1997, Mousseau2000, Mousseau2012} and the dimer method~\cite{Doye1997,Henkelman1999, Donghai2008} only use information related to the softest eigenmode of the Hessian. While the softest eigendirection in the dimer method is recovered by tracking and modifying the relative orientation of two neighbouring configurations of the PEL, ART evaluates the lowest eigenmode directly from the Lanczos algorithm~\cite{Lanczos1956}. In terms of efficiency, both methods compute the force multiple times at every iteration to optimize the dimer orientation and to generate a Krylov basis for the Lanczos algorithm respectively. 

From the literature, ART appears to be method of choice. It has been successfully applied in the following contexts: protein folding~\cite{Wei2003,Santini2004, StPierre2008, Derremaux2008}, PEL investigation of model polymers~\cite{Baiesi2009}, defects migration~\cite{Mellouhi2004, Machado2011} and structural excitations in amorphous materials~\cite{Mousseau1997, Ceresoli2006, Rodney2009, Kallel2010, Derlet2014, Derlet2016}. Despite the success of ART in sampling many relevant transition events, it remains a biased approach, as testified for instance in~\cite{Mousseau2000, Mousseau2012}, where the sample rate seems to decrease exponentially with the transition energy. The generality of this result is however unknown. On the other hand, for sufficiently small systems, completeness of the set of transitions can be claimed, allowing for reconstruction of the full dynamics~\cite{Mousseau2008, Mousseau2012}.

The development and implementation of saddle point search methods have been mostly devoted to structural systems, while investigations of magnetic systems are less common in the literature. In the work of Pastor and Jensen~\cite{Pastor2008}, transition states in a 2D disordered set of dipolar interacting spins are found by successively maximizing along the lowest initial eigenmode and relaxing in the perpendicular subspace. As opposed to ART, this simplified approach cannot inform about the neighbouring nature of the local minima, as one of its maximization-relaxation sequence may bypass all the neighbouring transition states. Adaptation of the NEB method was developed (geodesic NEB) and assessed on finding transition events leading to skyrmion and antivortex annihilation~\cite{Bessarab2015}. More recently, it was implemented to investigate the switching properties of elongated nanomagnets~\cite{Kevin2020}. On the other hand, ART has been adapted to study different transition events involving skyrmions~\cite{Mueller2018}. In general, for magnetic systems with fixed moment magnitudes, the non-Euclidean nature of the phase space requires special care. 

\begin{figure}[ht!]
    \includegraphics[width=\linewidth, trim=0cm 5.5cm 0cm 9cm, clip]{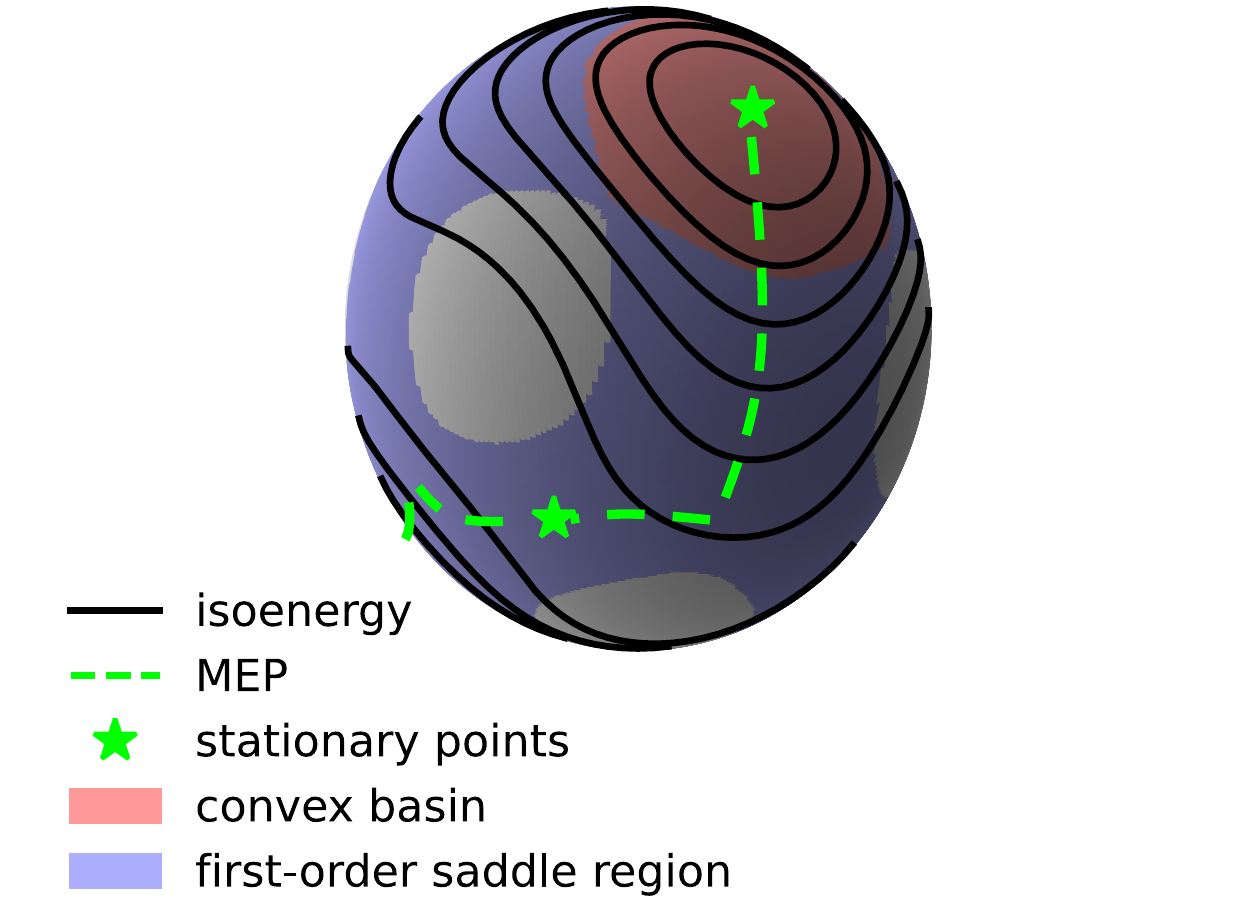}
\caption{\label{fig:MEL} Illustration of a Magnetic Energy Landscape (MEL) on a spherical manifold. Isoenergy lines correspond to the conservative Landau-Lifschitz dynamics. The minimum in the convex basin and the saddle point are connected by a minimum energy path (MEP), which lies at the bottom of the first-order saddle region which has exactly one negative Hessian mode. It is noted that the boundary of the convex basin does not correspond to an isoenergy line. The activation part of mART involves converging to a stationary (saddle) point within the first-order saddle region.} 
\end{figure}

Despite the relatively few implementations of saddle point search methods for magnetic energy landscapes (MEL), as in Fig.~\ref{fig:MEL}, compared to PEL, there is a priori a wider range of magnetic systems to which such methods could apply: due to the inherent presence of time reversal symmetry in the absence of external magnetic field, all configurations have a degenerate time reversal counterpart leading to canonical transitions to explore. For instance, while saddle point search methods are only applied on defected or disordered structural systems, a magnetic system with a translation invariant Hamiltonian and no applied field exhibits at least one relevant transition between its time reversal symmetric ground states. Saddle point search methods offer therefore a systematic approach to explore the magnetic switching mechanism.

The aim of this paper is to describe in detail the implementation of ART for magnetic systems with localized and continuous degrees of freedom, such as the classical XY or Heisenberg models. Such an implementation will be named mART, standing for magnetic ART. In addition, we formulate a criterion to efficiently recognize failed attempts and derive an expression for the step magnitude to control the convergence, irrespective of the investigated physical system. Finally, we also reveal interesting physical phenomena by implementing mART on two complex systems: a defected skyrmion lattice and a 2D dipolar spin glass.  

The paper is organized as follows: in section~\ref{sec:MEL}, we define the MEL and the relevant local properties. In section~\ref{sec:ART}, we describe mART as three parts corresponding to perturbing the initial minimum, finding a saddle point (activation) and relaxing to a neighbouring minimum (relaxation), and in section~\ref{sec:failure}, we discuss the criterion used to decide if a search should be halted and deemed a failure. In section~\ref{sec:stepMagnitude}, the derivation to obtain an adaptive step magnitude is presented. In section~\ref{sec:benchmarking}, we show the implementation of mART on two translational symmetric systems with isotropic exchange interactions, for which the outcome can be validated using known analytical results. In section~\ref{sec:skyrmions}, mART is used to study diffusion processes of typical 2D point defects in a skyrmion lattice. Finally, in section~\ref{sec:glass}, mART is run systematically to investigate activated transitions in a 2D dipolar spin glass. In section~\ref{sec:discussion}, we discuss the advantages and remaining difficulties of the present implementation and conclude on the physical relevance of the obtained results. 

\section{Method}
\subsection{\label{sec:MEL}Definition of the magnetic energy landscape and its derivatives}
In this work, we focus on a classical magnetic system consisting of $N$ spins $\bm{s}_{i}$, that are normalised $|\bm{s}_i|=1$. Due to this constraint, the manifold $\mathcal{M}$ constituting the phase space is non-Euclidean (Fig.~\ref{fig:MEL}). The Hamiltonian describing the MEL is defined for every configuration $m$ of $\mathcal{M}$: $\mathcal{H}(m): \mathcal{M} \rightarrow \mathbb{R}$. In general however, the constraint is dealt with separately and the Hamiltonian can be defined on the Euclidian space $\mathbb{R}^{3N}$ in which the manifold $\mathcal{M}$ is embedded: $\mathcal{\overline{H}}: \mathbb{R}^{3N} \rightarrow \mathbb{R}$. We assume a generic quadratic Hamiltonian on $\mathbb{R}^{3N}$: 
\begin{equation}
    \mathcal{\overline{H}}(\{\bm{s}_i\})=-\sum_i \bm{s}_i^T \bm{b}_i  + \frac{1}{2} \sum_{i,j} \bm{s}_i^T \mathbf{H}_{ij} \bm{s}_j \;,
\end{equation}
where $\bm{b}_i \in \mathbb{R}^{3}$ is a spin dependent external magnetic field and $\mathbf{H}_{ij} \in  \mathbb{R}^{3\times3}$ contains the pairwise interaction when $i\neq j$ (Heisenberg exchange or any type of anisotropic interactions such as dipolar or Dzyaloshinskii–Moriya interactions) or the single-ion anisotropy when $i=j$. For mART, it is required to evaluate locally the first and the second derivatives of $\mathcal{H}$ (see section~\ref{sec:ART}). To achieve this, the existence of the embedding space $\mathbb{R}^{3N}$ is handy. The gradient evaluated for a given configuration $m$ lives in the global tangent space, i.e. the union of the local tangent spaces of the spins $T_m \mathcal{M} = \bigcup_{i} T_{s_i} \mathcal{M}_i$, where $T_{s_i}\mathcal{M}_i=\{ \bm{\eta}_i \in \mathbb{R}^{3} : \bm{\eta}_i^T \bm{s}_i=0\}$. We can write it as
\begin{equation}
\begin{split}\label{eq:gradient}
    \mathrm{grad} \mathcal{H}(m)&=-\Motimes_{i} \mathbf{P}_{i}\left(\bm{b}_i - \sum_j \mathbf{H}_{ij} \bm{s}_j\right) \\
    &=-\Motimes_{i} \mathbf{P}_{i} \bm{h}_i\\
    &=-\Motimes_{i}\bm{h}_{\perp,i}\;, 
\end{split}
\end{equation}
where
\begin{equation}
\mathbf{P}_{i}=\left( \mathbf{I}_{3 \times 3} - \bm{s}_i \bm{s}_i^T\right)
\end{equation}
is a local projection operator,
\begin{equation}
    \bm{h}_i=\bm{b}_i - \sum_j \mathbf{H}_{ij} \bm{s}_j
\end{equation}
defines the local effective field and 
\begin{equation}
    \bm{h}_{\bot,i}= \mathbf{P}_{i} \bm{h}_i
\end{equation}
defines the component of the local effective field that is transverse to the spin. Of special interest for mART is the generalized force, i.e. the negative of the gradient, which, due to the previous definitions, can be identified as the global transverse magnetic field:
\begin{equation}
\begin{split}\label{eq:field}
    \bm{h}_{\bot}&=\Motimes_{i} \bm{h}_{\bot,i} \\
    &=\Motimes_i \left(\bm{b}_i-\sum_j \mathbf{H}_{ij} \bm{s}_j-\bm{s}_i \bm{s}_i^T \bm{h}_i  \right)\;.
\end{split}
\end{equation}
At this stage, we can define the stationary states, like the stable states (local minima) or transition states (first-order saddle points), as special configurations $m$, where the global transverse field $\bm{h}_{\bot}$ is zero. The Hessian is calculated as a regular directional derivative of $ \mathrm{grad} \mathcal{H}=-\bm{h}_{\bot}$ in $\mathbb{R}^{3}$ projected on the global tangent space, giving, for an arbitrary $\bm{\eta}=\bigotimes_i \bm{\eta}_i \in T_m \mathcal{M}$
\begin{equation}\label{eq:hessian1}
\begin{split}
    \mathrm{Hess} \mathcal{H}(m)[\bm{\eta}]&=\Motimes_{i} \mathbf{P}_{i}\left(\sum_j \mathbf{H}_{ij} \bm{\eta}_j + \bm{\eta}_i \bm{s}_i^T \bm{h}_i   \right) \\
    &= \Motimes_{i} \left(\sum_j \mathbf{P}_{i} \mathbf{H}_{ij} \bm{\eta}_j +  \bm{\eta}_i \bm{s}_i^T \bm{h}_i \right)  \;.
\end{split}
\end{equation}
In the first line above, the terms obtained from the derivation of $\bm{s}_i^T$ or $\bm{h}_i$ in Eqn.~\ref{eq:field} are not written since they will produce a $\mathbf{P}_{i}\bm{s}_i$ and therefore vanish. For numerical purposes, it remains to find a convenient basis of the global tangent space $T_m \mathcal{M}$ to express the Hessian. By choosing a basis where every vector is restricted to one local tangent space $T_{s_i}\mathcal{M}_i$, every entry of the Hessian can be written by the corresponding pair of spins. Such a basis presents two vectors for every $T_{s_i}\mathcal{M}_i$ that we label by $\mu$, $\nu \in \{1,2\}$, giving simply 
\begin{equation} \label{eq:hessian}
    \left< \bm{e}_{i \mu}, \mathrm{Hess} \mathcal{H}(m)[\bm{e}_{j \nu}]\right>=\bm{e}_{i \mu}^T \mathbf{H}_{ij} \bm{e}_{j \nu} + \delta_{\mu,\nu} \delta_{i,j} \bm{s}_i^T \bm{h}_i\;,
\end{equation}
 where $\left<\cdot, \cdot \right>$ denotes the scalar product in the global tangent space. A natural choice for the basis vectors $\bm{e}_{i1}$ and  $\bm{e}_{i2}$ are the two tangent spherical unit vectors spanning $T_{s_i}\mathcal{M}_i$, combined with two arbitrary perpendicular unit vectors at the poles. The first term in Eqn.~\ref{eq:hessian} comes from the Hessian of $\mathcal{\overline{H}}$, whereas the second term is a correction associated with the curvature of the manifold and is therefore constructed from the longitudinal derivative. Finally, we can distinguish between stable states and transition states by checking whether the lowest mode of the Hessian (Eqn.~\ref{eq:hessian}) is negative.

\subsection{\label{sec:ART}The mART algorithm}
Given a stable state of the MEL, mART proceeds in three subroutines to find a transition path towards a neighbouring stable state. The first and second subroutines compose the activation, while the third is the relaxation.

First, the exploration starts by perturbing the system away from the stable state. The conventional approach is to use a random direction of the phase space \cite{Mousseau2012}. Nonetheless, for systems with an ordered ground state, the low energy sector of the Hamiltonian contains configurations which break the order on the characteristic magnetic lengthscale, greatly constraining the directions for the search. This will be shown in the examples of sections~\ref{sec:benchmarking} and \ref{sec:skyrmions}. Alternatively, information about the transitions can be contained in the harmonic modes, which are retrieved from diagonalizing the Hessian (Eqn.\ref{eq:hessian}) at the initial stable state. This possibility will be shown in section~\ref{sec:spinChain} and \ref{sec:glass}. Once the perturbation is chosen, it can be followed all the way until the lowest eigenvalue of the Hessian turns negative, a configuration which defines the boundary of the convex basin associated with the initial state (Fig.~\ref{fig:MEL}).
 
Second, to proceed away from the convex basin, the progression direction $\bm{g}$ is iteratively selected to relax the system along the global transverse field $\bm{h_{\bot}}$, except in the direction of the lowest Hessian mode $\bm{q}_0$: 
\begin{equation}
\begin{split}
\bm{g} &\propto \bm{h}_{\bot}-(1+\gamma)\left< \bm{h}_{\bot}, \bm{q}_0 \right> \bm{q}_0 \\
&\propto \sum_{n\geq 1} \left< \bm{h}_{\bot}, \bm{q}_n \right> \bm{q}_n - \gamma  \left< \bm{h}_{\bot}, \bm{q}_0 \right>\bm{q}_0
 \label{eq:progression}
 \end{split}
\end{equation}
with the second equation obtained by expanding in the basis of the Hessian modes, showing the relative weight attributed to the optimization in each basis direction. $\gamma$ is a positive parameter that controls the maximization along $\bm{q}_0$ relative to the minimization in the transverse hyperplane. A natural choice for $\gamma$ is one, which means that all directions are treated on equal footing with respect to the generalized force, as in the optimization during the gradient descent algorithm. For all the systems studied in this paper, this turns out to be a good choice. How to chose the magnitude of the step along $\bm{g}$ is given later by Eqn.~\ref{eq:alpha}. The progression direction $\bm{g}$ is a vector of the global tangent space $T_m \mathcal{M}$ and a retraction to the manifold $\mathcal{M}$ is necessary at the end of every iteration. The retraction consists simply of renormalising every spin. The transition state is formally identified when the norm of $\bm{h_{\bot}}$ equals zero. In practice it is found when this quantity reaches a preset minimum value. 

Third, the search for a neighbouring stable state continues by pushing the transition state configuration a constant step away from the initial state. From there, any minimization algorithm can be used. For simplicity, we choose the gradient descent algorithm.

The second step of mART relies on iteratively computing the lowest eigenvalue and corresponding eigenvector of the Hessian. To achieve this, the Hessian can be computed entirely in $\mathcal{O}(N^2)$ operations, with $N$ the number of spins, thanks to the existing closed-form expression (Eqn.~\ref{eq:hessian}) and with its lowest eigenpair found efficiently by the Lanczos algorithm~\cite{Lanczos1956}. The Lanczos algorithm runs in $\mathcal{O}(pN^2)$, with $p$ the dimension of the Krylov basis. By using the eigenvector obtained at the previous iteration as the first vector of the Krylov basis, we observe a rapid convergence for the lowest eigenpair, such that $p$ is selected to never exceed 30 in all our simulations. In other examples~\cite{Mousseau2012}, convergence was obtained for $p$ as small as 15, independent of system size.  

\subsection{\label{sec:failure}Recognizing a failed search}
We consider saddle points that are adjacent to the initial minimum, as those that can be connected by a monotonously increasing energy path. Such an arbitrary transition path can always be relaxed in its transverse hyperplane to give a minimum energy path (MEP) which can be seen as an ensemble of configurations $m$ such that $\left<\bm{h}_{\bot}(m), \bm{q}_0(m)\right>=\sqrt{\left<\bm{h}_{\bot}, \bm{h}_{\bot}\right>}$ (Fig.~\ref{fig:MEL}). Beyond the convex basin, an MEP lays at the bottom of a region with one negative Hessian mode, a first-order saddle region (Fig.~\ref{fig:MEL}). In contrast, due to the relaxation transverse to the lowest mode, mART exploration path is not guaranteed to be a monotonously increasing energy path and, therefore, the adjacency of the found saddle point cannot be inferred. Nevertheless, the found saddle point is almost certainly adjacent to the initial minimum if the search proceeds in the first-order saddle region, which is composed of nearly disconnected valleys surrounding the MEPs as in Fig.~\ref{fig:MEL}. On the contrary, if the search extends to regions with two or more negative eigenvalues, it can easily skip adjacent saddle points and converge towards non-adjacent ones. Consequently, we require for the second eigenvalue to stay positive during the whole exploration, otherwise the attempt is discarded. Fortunately, the second lowest eigenvalue of the Hessian is sufficiently well approximated by the Lanczos algorithm without expanding the Krylov basis. In addition, the proximity of the exploration path to an MEP can be assessed during the search by monitoring the quantity $\left<\bm{q}_0, \bm{h}_{\bot}\right>/\sqrt{\left<\bm{h}_{\bot}, \bm{h}_{\bot}\right>}$, which should remain close to one.

Under such conditions, failures are attributed to an inappropriate choice of the perturbation which sets the search too far away from an MEP and therefore do not lead to a transition state. Practically, the exploration rapidly exits the first-order saddle region after the perturbation. Nonetheless, one may correct for an inappropriate perturbation by increasing the relaxation in the transverse hyperplane of the lowest mode (reducing $\gamma$ to a value smaller than one). One detrimental effect of this choice might be to set back the exploration at the initial minimum, where the deterministic progression in the second step of the activation can only find a unique saddle point.

Since the gradient descent algorithm initialized at a saddle point necessarily connects to an adjacent minimum, as it reduces the energy at every step, one may ultimately confirm adjacency if the gradient descent is able to recover the initial minimum. Notwithstanding, a saddle point can be adjacent to more than two minima and the gradient descent initialized at different locations in the vicinity of the saddle point might only be able to find a few of them, excluding the initial minimum.

\subsection{\label{sec:stepMagnitude}Adaptive step magnitude}
Even for a quadratic Hamiltonian $\mathcal{\overline{H}}$ in the $\mathbb{R}^{3N}$ embedding space, the corresponding MEL $\mathcal{H}$ on $\mathcal{M}$ is intrinsically non-harmonic. It is therefore not straightforward to find a step magnitude along $\bm{g}$ or $\bm{h}_{\bot}$, respectively in the second and third subroutines of the algorithm, which automatically prevents the search from missing the stationary points ($\bm{h}_{\bot}=\bm{0}$). We propose a scheme to evaluate the step magnitude that relies on finding a trust distance for the first order approximation of the MEL. Since the quadratic expansion is readily available at every iteration, it can be used to approximate the actual energy change and avoid an arbitrary update of the step magnitude based only on the prior iteration. For a step magnitude $\alpha \ll 1$ along the normalised progression vector $\bm{\hat{g}}=\bm{g}/\sqrt{\left<\bm{g}, \bm{g} \right>}$, we evaluate a trust ratio as the ratio between the first and the second order expansion of the MEL:
\begin{equation}
\rho=\frac{-\alpha \left<\bm{\hat{g}}, \bm{h}_{\bot}\right>}{-\alpha \left<\bm{\hat{g}}, \bm{h}_{\bot}\right> +\alpha^2  \left< \bm{\hat{g}}, \mathrm{Hess} \mathcal{H}(s)[\bm{\hat{g}}]\right>} \;. \label{eq:trustRegion}
\end{equation}
If the trust ratio is close to one, the change in the MEL due to the curvature is small from one iteration to the next and therefore the change in the transverse field $\bm{h}_{\bot}$ is controlled, such that stationary points cannot be overshot. In practice, the trust ratio may be bound between $1-\epsilon$ and $1+\epsilon$, where $\epsilon \ll 1$ is a preset positive control parameter. This allows to solve for the positive step magnitude $\alpha$:
\begin{equation}
    \alpha=\min\left(2\epsilon \left|\frac{\left<\bm{\hat{g}}, \bm{h}_{\bot}\right>}{\left< \bm{\hat{g}}, \mathrm{Hess} \mathcal{H}(s)[\bm{\hat{g}}]\right>} \right|,0.1\right)\;, 
\end{equation}
where the upper bound guarantees the validity of the expansions in Eqn.~\ref{eq:trustRegion}. This expression produces the following behaviour: a large step magnitude due to the denominator between the stationary points as the search progresses along directions of low or no curvature, like the inflection point at the convex basin boundary, and a small step magnitude at the approach of the stationary points due to the numerator. Nevertheless, the numerator presents also solutions when $\bm{g}$ is perpendicular to $\bm{h}_{\bot}$, creating other locations where the search slows down. To avoid this, we rewrite the progression magnitude assuming $\bm{g}$ and $\bm{h}_{\bot}$ are colinear in the numerator, which corresponds to the exploration path following an MEP:  
\begin{equation}
    \alpha=\min\left(\frac{2\epsilon \sqrt{\left<\bm{h}_{\bot}, \bm{h}_{\bot}\right>}}{\left|\left< \bm{\hat{g}}, \mathrm{Hess} \mathcal{H}(s)[\bm{\hat{g}}]\right>\right|},0.1\right)\;. 
\label{eq:alpha}
\end{equation}
The progression magnitude as computed from Eqn.~\ref{eq:alpha} allows for a controlled convergence to the stationary points, while minimizing the number of iterations.
 
\section{\label{sec:benchmarking}Validation on known systems}
\subsection{\label{sec:spinChain}Periodic spin chains}

\begin{figure*}[ht]
    \includegraphics[width=\linewidth, trim=0cm 6.8cm 0cm 7cm, clip]{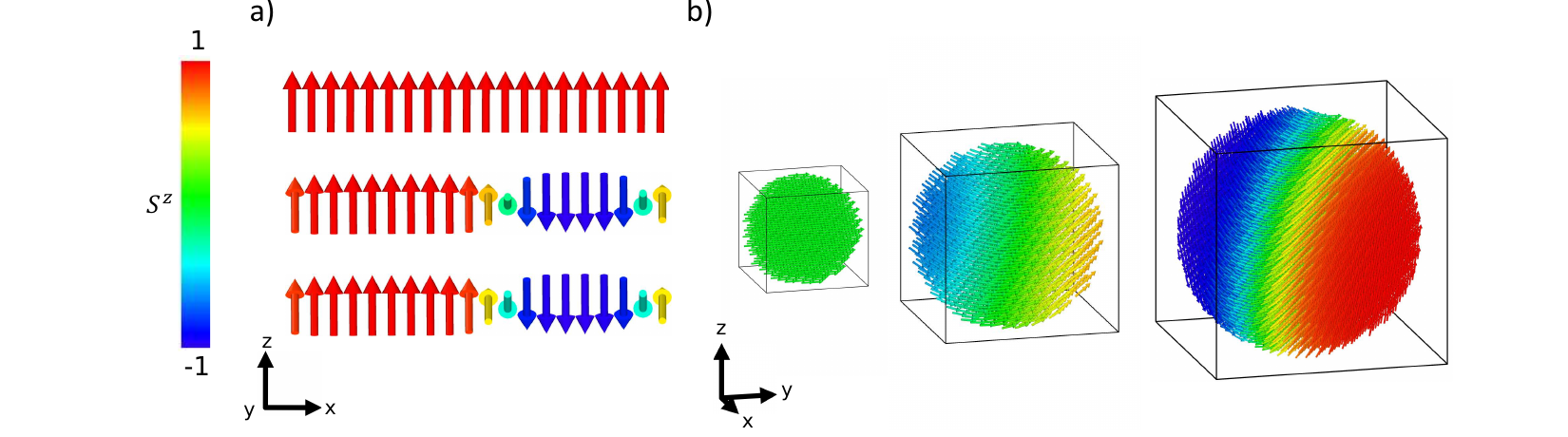}
\caption{\label{fig:spinChainFigure} a) (upper panel) Initial ground state configuration for an $N=20$ spin chain ferromagnetically ordered along +$z$. (middle panel) Configuration of an adjacent transition state found by mART including a domain with opposite magnetization ($N_m=\sqrt{J/K_z}=\sqrt{3}$, corresponding to $N \sim N_m$). (lower panel) Configuration of a neighbouring stable state. b) Transition states between the two ferromagnetically ordered ground states of ideal HCP ($a=b$, $c=\sqrt{8/3}a$) nanoparticles with $N_m=\sqrt{10}$. The particle diameters are respectively $10.1a$, $12.8a$ and $22.3a$.} 
\end{figure*}

In this first case, we consider a periodic spin chain with uniform ferromagnetic exchange and anisotropy. We choose to look at this system, because it is a simple example of a magnetic system with a lattice translation invariance and two degenerate ground states due to the time reversal symmetry. The transition path between these ground states can be investigated with mART. The system is composed of spins located along the x-axis with periodic boundary conditions. The expression for the MEL is given by the following Hamiltonian: 
\begin{equation}
    \mathcal{\overline{H}}= -J \sum_{<i,j>} \bm{s}_i \cdot \bm{s}_j - K_z\sum_i (s_i^z)^2\;,
\end{equation}
where $J$ is the exchange constant between nearest neighbours and $K_z$ an easy z-axis anisotropy. The parameters are positive and given in units of energy. The system admits metastable states with magnetically reversed domains in addition to the two Ising-like ground states with ferromagnetic order along +z and -z. The domains walls at the interfaces of the magnetic domains scale with the characteristic magnetic length $N_m=\sqrt{J/K_z}$, when the latter is much greater than one, i.e. in the continuum spin regime~\cite{Landau1935}. Without loss of generality, we add a strong easy-plane anisotropy to favour Bloch rather than Néel walls.
 
We apply mART to find a transition path between the Ising-like ground states. To initialize the procedure, a perturbation must be chosen. Instead of exploring along random directions, the ground state can be perturbed along the harmonic modes, which constitute spin wave modes commensurate with the system size. However, since these are delocalised excitations, they project badly on the expected bound transition states. As a consequence, mART fails to recover transition states, except in two regimes: first, when $N_m \gg N$, the harmonic mode with an infinite wavelength (zero-mode) finds the uniform rotation as a transition state. Second, when $N_m \sim N$, the harmonic modes with wavelength comparable to the system size find a transition involving a reversed domain that spans approximately half of the system (Fig.~\ref{fig:spinChainFigure}a).

\begin{figure}[h]
    \includegraphics[width=1.0\linewidth,trim=0.5cm 0.5cm 0.3cm 0.4cm, clip]{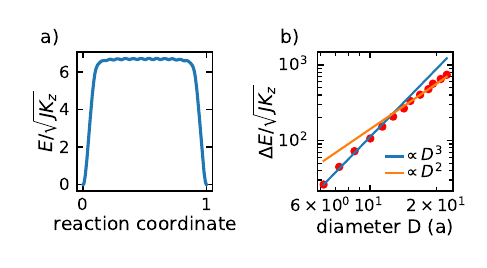}
     \caption{\label{fig:spinChainPlots} a) Transition path between the two ferromagnetically ordered ground states of an $N=20$ spins chain ($N_m=\sqrt{J/K_z}=\sqrt{2}$) found by repeatedly applying mART. b) Energy barrier between the two ferromagnetically ordered ground states of an ideal HCP ($a=b$, $c=\sqrt{8/3}a$) nanoparticle with respect to its diameter ($N_m=\sqrt{10}$). The two polynomial curves are guides to the eye.}
\end{figure}

Alternatively, the perturbation can be selected from an educated guess relying on the fact that the low energy sector of the Hamiltonian contains configurations which break the translational order only on the characteristic magnetic lengthscale. The perturbation from the ground state can thus act uniformly on a number of spins of the order of $N_m$. For sufficiently large systems ($N_m \ll N$), this allows to systematically recover a transition to a metastable state containing a small reversed magnetic domain, i.e. a magnetic defect. In addition, for $N_m > 1$, the defect energy is found to be approximately proportional to the characteristic magnetic energy $E_m=\sqrt{JK_z}$ (not shown). This is theoretically expected and validates our method, since defects are made of two domain walls whose energies scale as $E_m$ in the continuum spin regime $N_m \gg 1$. Once the first metastable state is reached, the procedure can be restarted to find the next metastable state. This is done by using a perturbation whose direction is opposite to the direction pointing towards the initial state. This procedure can be repeated to unveil the entire transition path between the two ground states (Fig.~\ref{fig:spinChainPlots}a). Once nucleated, the magnetic defect expands through the propagation of its domain walls until it merges with its periodic image. We note that the energy barriers for the expansion of the defect are small compared to the nucleation energy, which is of the order of $E_m$. 

When approaching the continuum spin regime $N_m\gg1$, the expansion energies become smaller, the number of metastable states reduces and the size of the defects gets bigger. mART takes gradually more iterations to find the transition states, as the MEL flattens close to them reducing significantly the step magnitude (Eqn.~\ref{eq:alpha}). This is consistent with the fact that in the limit $N_m \rightarrow \infty$, the continuous degeneracy of the internal SO(3) symmetry of the exchange is recovered, making the search for transitions irrelevant. 

\subsection{HCP nanoparticles}\label{sec:nanoparticles}
We focus in this section on a spherical magnetic nanoparticle. The crystal structure of the particle is ideal hexagonal close-packed (HCP), characterised by unit cell vectors $a=b$, $c=\sqrt{8/3}a$, such that every spin has 12 nearest neighbours (six in the same stacking planes and three in every neighbouring plane). The exchange interaction $J$ is between the nearest neighbour pairs. The crystal point symmetries admit a quadratic anisotropy term along $\left<001\right>$. Consequently, we choose an easy-axis anisotropy $K_z$ along $\left<001\right>$ coinciding with the z-axis of our geometry.

The transition state separating the two ferromagnetically ordered ground states depends on the particle diameter. Recalling the asymptotic behaviour discussed in section~\ref{sec:spinChain}, for a particle with a small diameter with respect to the characteristic magnetic length $N_m=\sqrt{J/K_z}$, we expect a uniform rotation of the magnetization for the lowest energy transition between the ground states, such that the energy barrier scales with the volume ($\propto D^{3}$). In contrast, for a large particle, it is favourable to frustrate the exchange interaction and create a domain wall, leading to an energy barrier scaling with the great disc ($\propto D^{2}$), where the domain wall interface is the largest. We want to retrieve the crossover between these regimes with mART and investigate the transition configurations in the regime of domain walls. 

We set $N_m=\sqrt{10}$ and chose particles with diameters ranging from $6.2a$ to $22.3a$. For all simulations, the initial ground state is perturbed by an homogeneous rotation about the x-axis. The resulting transition states highlighting the two regimes and the crossover are presented in Fig.~\ref{fig:spinChainFigure}b. For the largest particles, it is relevant to notice the resilience of the implementation to retrieve non-homogeneous magnetic configurations by starting with a perfectly homogeneous perturbation. In more detail, for particles of diameter equal or greater than $12.8a$, mART gives transition states with three different domain wall orientations, which correspond to the three prismatic planes ($\{10\bar{1}0\}$). This result can be justified by the low density of bonds in the prismatic planes of the ideal HCP structure ($\rho_{\{10\bar{1}0\}}=\sqrt{6}/a^2$), which is lower than the density of bonds in the basal plane ($\rho_{(0001)}=6/(\sqrt{3}a^2)$). Therefore, the domain wall develops preferably in the three prismatic planes and mART recovers primarily those transitions. This results in a 6-fold degeneracy for the lowest energy transitions due to the presence of the three equivalent prismatic planes and the odd parity for the magnetization about the domain walls. We can notice however a small misorientation between the domain wall planes and the respective prismatic planes. This is attributed to the roughness of the boundaries: spins at the boundary have different coordination numbers due to the discretisation of the sphere, allowing for some regions at the surface to reduce the exchange energy of the wall. We can finally verify that the expected scaling of the energy barrier is recovered, by plotting it with respect to the particle diameter (Fig.~\ref{fig:spinChainPlots}b).

\section{Transitions in an emergent skyrmion lattice}\label{sec:skyrmions}
\begin{figure}[h]
    \includegraphics[width=\linewidth, trim=0.1cm 1cm 0.1cm 1.5cm, clip]{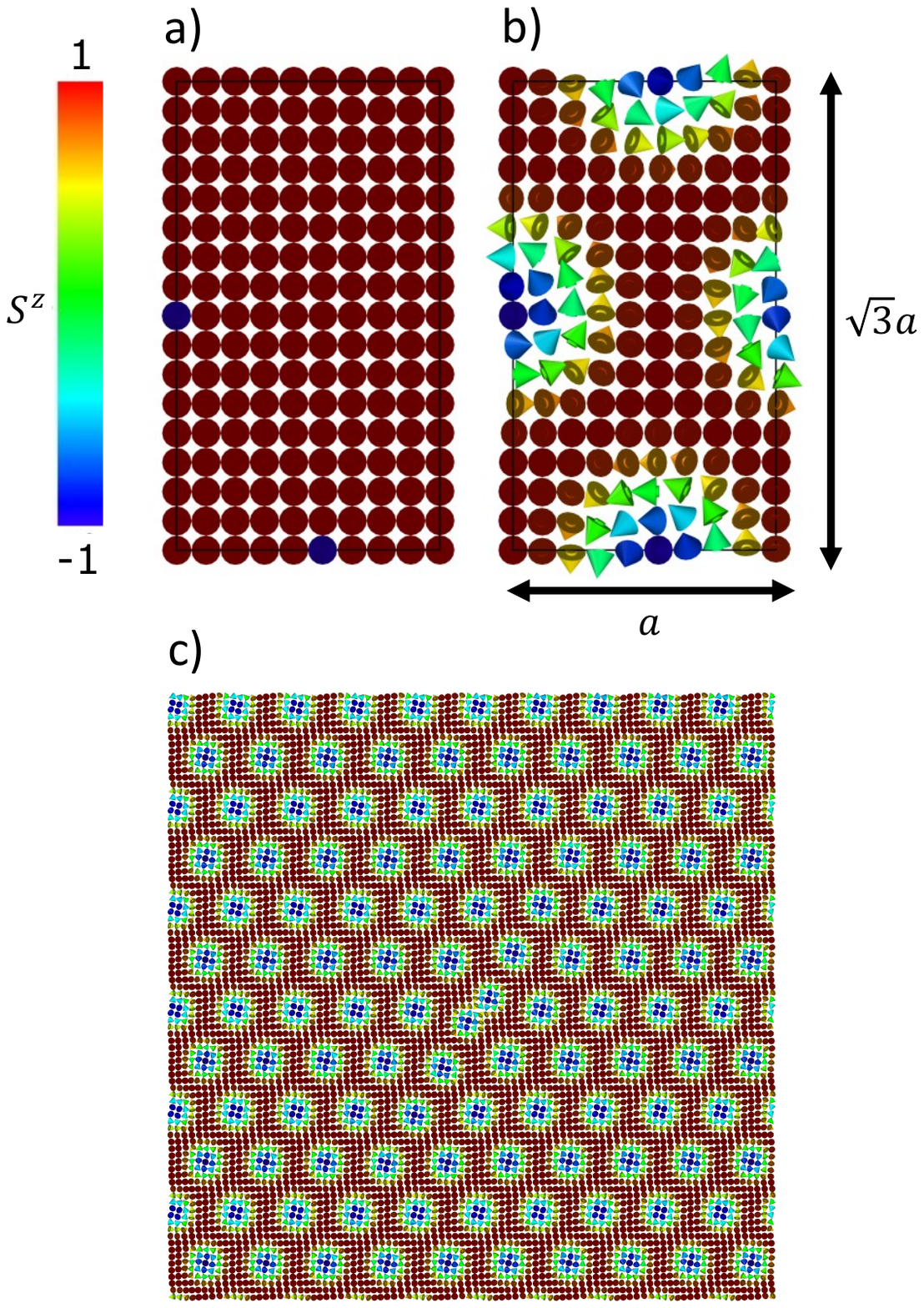}
     \caption{\label{fig:skyrmionUnitCell} a) Initial configuration used to find the equilibrium lattice parameter of the hexagonal lattice in the skyrmion phase (SkX). Two spins of the ferromagnetic arrangement are flipped. The aspect ratio of the periodic simulation cell is chosen to represent one unit cell of the hexagonal SkX lattice. b) The same unit cell after relaxation with the gradient descent algorithm. The procedure repeated for different unit cell sizes yields a minimal relaxed energy used to define the equilibrium lattice parameter. c) Skyrmion lattice configuration hosting a defect at the center. This corresponds to a transition state configuration of the interstitial defect.}   
\end{figure}
In this section, we aim at using mART on a more complex system composed of interacting skyrmions, for which theoretical predictions are limited. A skyrmion is a type of 2D magnetic texture with local inversion of the magnetization, and whose stability can be understood from its non-trivial topology. It emerges for instance as a metastable solution of the simplest isotropic exchange Hamiltonian~\cite{Polyakov1975}. The presence of skyrmions becomes statistically relevant when the exchange is anisotropic and contributes to a Dzyaloshinskii–Moriya interaction (DMI) term. We consider the following nearest neighbour Hamiltonian on a square spin lattice:   
\begin{equation}\label{eq:hamiltonianSkyrmions}
    \mathcal{\overline{H}}= - B\sum_i s_i^z  - J \sum_{<i,j>} \bm{s}_i \cdot \bm{s}_j +  D \sum_{<i,j>} \bm{r}_{ij} \cdot (\bm{s}_i \times \bm{s}_j) \;,
\end{equation}
where $B$ is a magnetic field, $D$ is the DMI constant and $\bm{r}_{ij}$ is the displacement vector between spin $i$ and $j$. The interactions are restricted to the nearest neighbours. The DMI interaction breaks the inversion symmetry between two sites and stabilizes chiral textures such as spirals and SO(2) symmetric Bloch skyrmions (Fig.~\ref{fig:skyrmionUnitCell}b). The magnetic field $B$ breaks the time reversal symmetry disfavouring the non-magnetic spiral texture. This system supports therefore a ground state skyrmion phase (SkX), characterised by an emergent hexagonal skyrmion lattice, when the domain wall width $N_D=J/D$ is comparable to the magnetic screening length $N_H=\sqrt{J/H}$. More precisely, the condition is $0.45 \leq N_D/N_H \leq 0.90$~\cite{Banerjee2014, Utkan2016}. The ground state hexagonal lattice parameter is difficult to predict as the distance dependence in the interaction can only be approximated from a numerical investigation~\cite{Capic2020}. The available analytical solution in the pure exchange limit depends only on the topological charge and is distance independent.

\begin{figure*}[ht]
\includegraphics[width=0.7\linewidth, trim=1cm 5cm 1cm 3cm, clip]{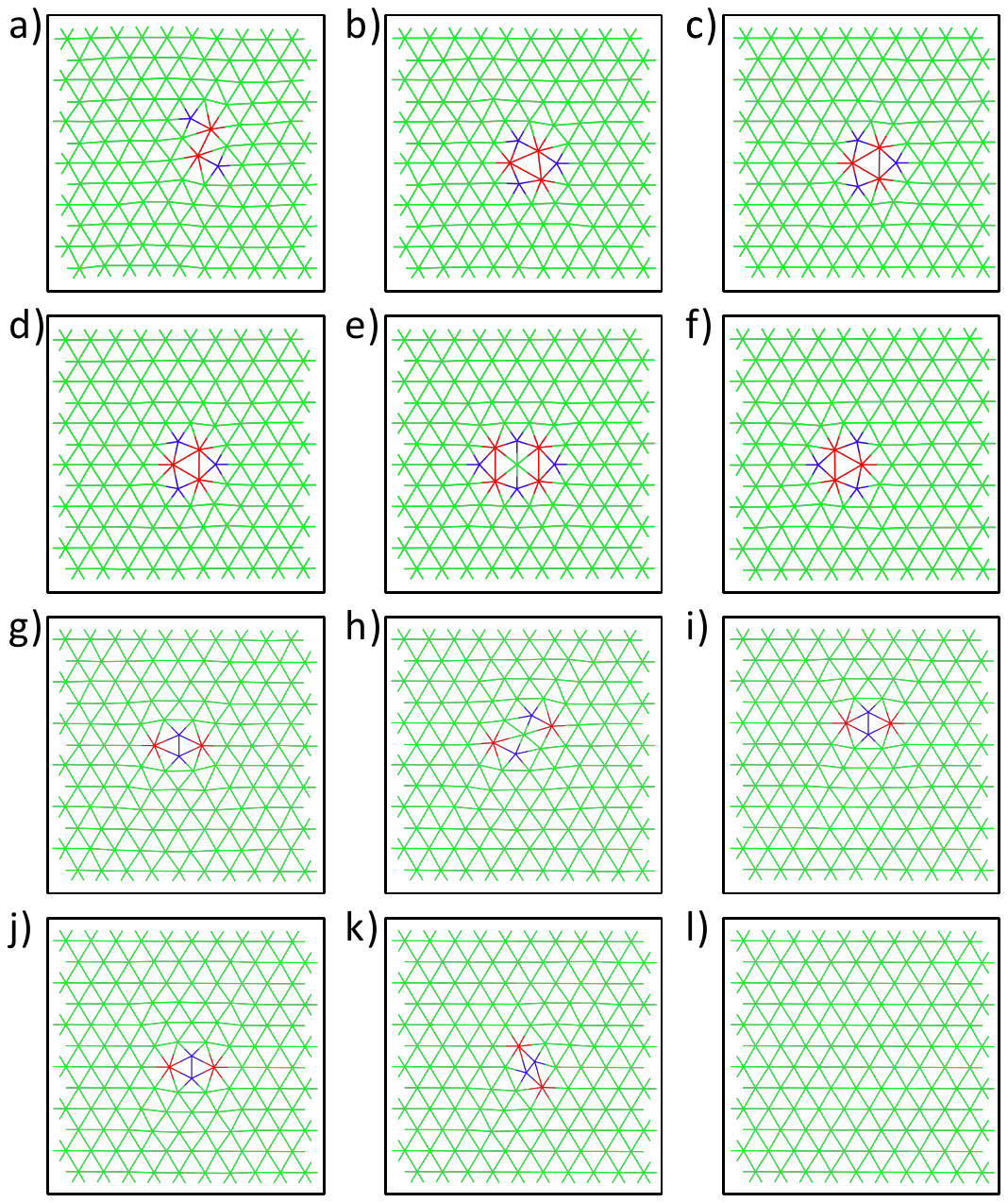}
     \caption{\label{fig:skyrmions} a) Initial stable skyrmion lattice with a two-fold symmetric vacancy represented by its bonding structure where each skyrmion is treated as a point object. Five-, six- and seven-fold coordinations at the vortices are colored blue, green and red respectively. b) Transition state configuration found by mART with an energy barrier $\Delta E=0.18J$. c) Neighbouring stable state configuration corresponding to a three-fold symmetric vacancy and the corresponding negative energy barrier from the transition state is $\Delta E'=0.05J$. d-f) With the same logic, the transition process representing the reflection of a three-fold symmetric vacancy along one of its edges ($\Delta E=\Delta E'=0.05J$). g-i) The transition process for a two-fold symmetric interstitial corresponding to a gliding along one of the lattice direction ($\Delta E=\Delta E'=0.12J$). j-l) The transition process for a two-fold symmetric interstitial corresponding to annihilation by merging two skyrmions ($\Delta E=2.18J$ and $\Delta E'=4.65J$). The spin configuration at the transition is shown in Fig.~\ref{fig:skyrmionUnitCell}c.}
\end{figure*}

At low enough temperature and due to their topological nature, skyrmions are expected to primarily interact at distance rather than create or annihilate, which makes them behave like a particle ensemble. The underlying square spin lattice, which break the six-fold symmetry of the hexagonal skyrmion lattice, does not provide the commensurate effective periodic potential required for the low temperature phase of this 2D system to exhibit translational order~\cite{Nelson_1979}. On the contrary, in the continuum spin regime, when the skyrmion diameters are large compared to the spin lattice parameter ($N_H\sim N_D\gg1$), the skyrmion system must exhibit the same point defects that break down the order in particle systems at low temperature, i.e. interstitials and vacancies~\cite{Kosterlitz_1973, Nelson_1978}. Due to their topological nature, skyrmion interstitials and vacancies are metastable, which allows for investigation by mART. To keep the problem computationally tractable, while aiming at observing defect behaviours reminiscent of particle systems, we set $N_D=1$ for $N_D/N_H=0.71$. 

The ground state lattice parameter is found as follows: a ferromagnetically ordered configuration with two distant spins flipped is set up in a rectangular simulation cell with an aspect ratio of $\sqrt{3}$ under periodic boundary conditions (Fig.~\ref{fig:skyrmionUnitCell}a). For a range of simulation cell sizes, the configuration is relaxed (Fig.~\ref{fig:skyrmionUnitCell}b) and the final energy is computed to obtain an equation-of-state curve, the minimum energy of which gives the equilibrium skyrmion lattice parameter. We then construct a  crystal skyrmion lattice using this lattice parameter, by relaxing a system of $10\times6$ ferromagnetic unit cells (8100 spins), each containing two flipped spins. The defected lattices, starting configurations for mART, are created by adding or removing a flipped spin before the relaxation. 

Applying mART in the proposed way turns out to be problematic for the chosen parametrisation ($N_H\sim N_D\sim1$), as the energy of the global translational modes, despite of being finite compared to the case $N_H\sim N_D\rightarrow \infty$, is small. At the beginning of the second subroutine of mART, while the search falls back inside the convex basin for several steps due to the relaxation, one of these modes, having the smallest energy, is followed. Nonetheless, in contrast to the spin chain in section~\ref{sec:spinChain} with $N_m\sim1$, the transitions between the degenerate translational symmetric configurations resulting from these modes are not of interest here. Therefore, we apply the perturbation further beyond the convex basin boundary to single out a more specific transition. In particular, by displacing a skyrmion in the region of the defect until the second lowest mode becomes negative, the search successfully identifies a localized transition path.

We are primarily interested in diffusion processes of vacancy and interstitial defects. Starting with the two-fold symmetric vacancy in Fig.~\ref{fig:skyrmions}a, a transition constituting a central reflection is found with an energy barrier of $\Delta E=0.30J$ (not shown). While the latter does not allow for the vacancy to move, the transition in Fig.~\ref{fig:skyrmions}a-c corresponds to an excitation with an energy barrier $\Delta E=0.18J$ into a higher energy stable three-fold symmetric vacancy which can diffuse according to the process in Fig.~\ref{fig:skyrmions}d-f ($\Delta E=0.05J$). Overall, the two-fold symmetric vacancy can diffuse through a three-step process, which includes two intermediate states composed of three-fold symmetric vacancies (Fig~\ref{fig:skyrmions}d and \ref{fig:skyrmions}f). We then search for an activated diffusion process for a two-fold symmetric interstitial. By first noting that an isolated pair of five- and seven-fold neighbouring sites form a dislocation, we observe a diffusion process in Fig.~\ref{fig:skyrmions}g-i during which the pair of dislocations forming the interstitial unbind, requiring an energy of $\Delta E=0.12J$. Finally, we retrieve a process by which the interstitial annihilates (Fig.~\ref{fig:skyrmions}j-l), quantifying therefore the topological protection discussed above which guarantees the mapping to a particle system: it requires an energy of $\Delta E=2.18J$ to annihilate the defect and $\Delta E'=4.65J$ to create it on the ground state, which are both an order of magnitude larger than the energies of the other processes.

The vacancies in Fig.~\ref{fig:skyrmions}a and in Fig.~\ref{fig:skyrmions}d and the interstitial in Fig.~\ref{fig:skyrmions}g are statistically relevant defects of 2D particle systems~\cite{Libal_2007, Kim_2020}. On the other hand, other stable defects of 2D systems, in particular the axial two-fold symmetric vacancy and the three-fold symmetric interstitial, are not found to be stable. Similarly, we note that the split interstitial (Fig.~\ref{fig:skyrmions}h) appears as a saddle point, in contrast to the observations in~\cite{Libal_2007, Kim_2020}. These results can be seen as the effect of the underlying square spin lattice, which remains present for finite values of $N_H\sim N_D$. Here, it does not only affect the energy levels of the stable states, but can also modify the stability of the states, as we confirm by observing that the interstitial in Fig.~\ref{fig:skyrmions}g is not stable along the two other skyrmion lattice directions. Overall, this reduces the number of possible transition paths and no other type of diffusion processes were found with mART by displacing skyrmions in the neighbourhood of the present common defects. We finally note that the computed diffusion process for the most stable vacancy requires an energy one and half time higher than the one for the interstitial. This predicts a faster diffusion for the interstitial compared to the vacancy, in agreement with results in particle systems~\cite{Libal_2007, Kim_2020}.

\section{Barrier energy statistics of a 2D dipolar spin glass}\label{sec:glass}

\begin{figure*}[ht!]
\includegraphics[width=1\linewidth, trim=3cm 5cm 3cm 7.5cm, clip]{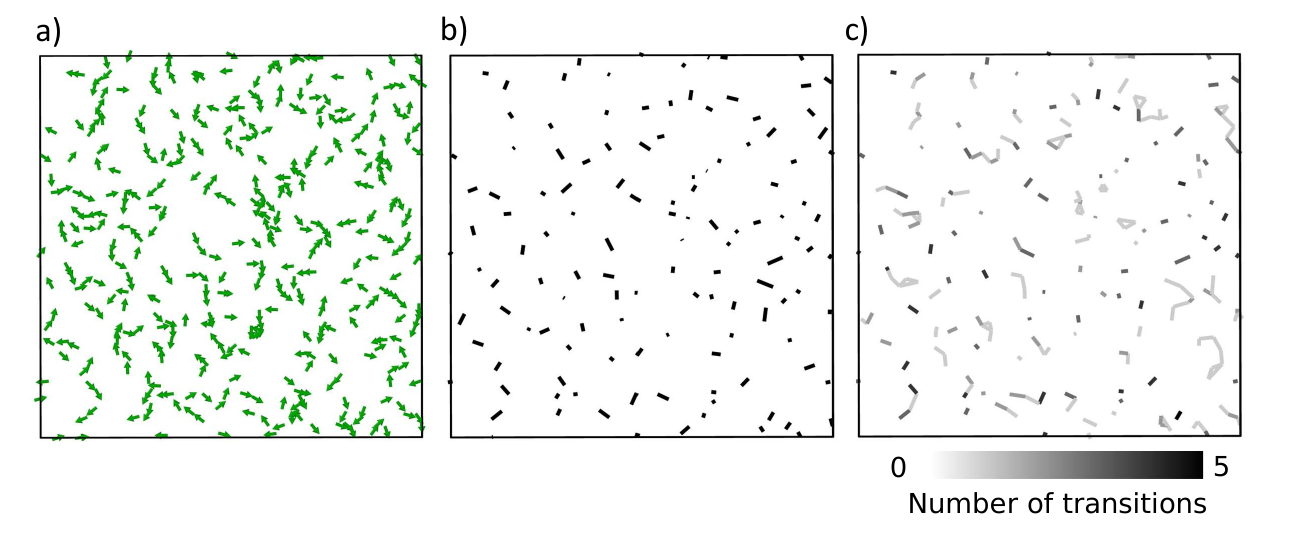}
\caption{\label{fig:spinGlass} a) Initial metastable configuration of the 2D dipolar glass. b) Bonds between nearest neighbour pairs (NNP)s, i.e. pairs of spins that are mutually nearest neighbours. c) Bonds between the most contributing pairs of each transition (Eqn.~\ref{eq:mostContributing}), colored according to the number of occurrences.}
\end{figure*}

mART, as an open-ended technique to find transitions, can be used to investigate the large number of transitions in a disordered system, giving insights into activated dynamics that might be difficult to predict analytically. The presently studied structure is composed of in-plane spins located in a 2D square according to a Poisson point process, i.e. the spatial positions of the spins are independently and uniformly distributed in space (Fig.~\ref{fig:spinGlass}a). The spins interact with each other via the dipolar interaction:

\begin{equation}\label{eq:hamiltonianDipolar}    \mathcal{\overline{H}}= \frac{\mu_0}{4 \pi}  \sum_{i>j} \frac{  \bm{s}_i \cdot \bm{s}_j - \left(\bm{s}_i \cdot \bm{\hat{r}}_{ij}\right) \left(\bm{s}_j \cdot \bm{\hat{r}}_{ij}\right) }{|r_{ij}|^3}\;,
\end{equation}
where $\mu_0$ is the vacuum permeability. 
Due to the explicit quenched disorder in the Hamiltonian inherited from the positional disorder, this system is a spin glass. Nonetheless, due to the short-range nature of the interaction, it is not a mean-field model in the thermodynamic limit, implying that ergodicity does not break down~\cite{Castellani_2005}, in contrast for instance to the site-diluted dipolar Ising system in 3D~\cite{Alonso_2010}. Practical consequences of this observation are that we can start our analysis with a random magnetic configuration relaxed to the first stable state in a system with periodic boundary conditions and a cutoff in the interaction. In particular, we consider a system of 400 spins with cutoff in the interaction distance of $5\rho^{-0.5}$, where $\rho$ is the 2D density of spins.

\begin{figure}[h]
\includegraphics[width=1.0\linewidth,trim=0.4cm 0.4cm 0.3cm 0.3cm, clip]{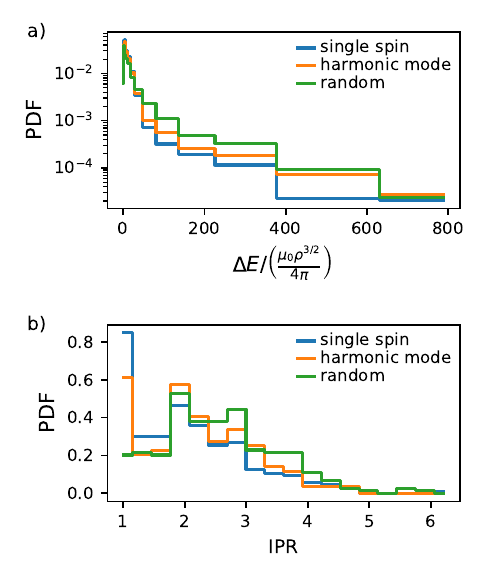}
\caption{\label{fig:IPRAndEnergyBarriers} a) Log-linear plot of the distributions of normalised energy barriers obtained for the 284 different transitions found with single spin perturbations, the 349 with harmonic mode perturbations and the 243 with random perturbations. b) Distributions of the inverse participation ratio (IPR) according to the type of perturbation used to find the transition state.}
\end{figure}

The quenched disorder combined with the anisotropic interaction discard the presence of a continuous degeneracy, making stable and transition states well defined~\cite{Belobrov1982,Zimmerman1988}. We first perturb individually every spin with a rotation in one or the other direction. This gives 800 different initial starting directions for mART which finally result in 281 different transition states. Fig.~\ref{fig:IPRAndEnergyBarriers}a plots the distribution of energy barriers $\Delta E$ normalised by the energy scale, that is $\mu_0 \rho^{3/2}/4\pi$ (blue curve). The obtained distribution appears to asymptote to zero as the energy approaches zero (as expected by the absence of soft modes), is peaked at about the energy scale and exhibits a heavy tail. To characterise these transition states, we measure their degree of localization via the inverse participation ratio (IPR), computed from the individual changes of angles between the initial state and the transition state $\Delta \phi_i$: 
\begin{equation}
\begin{split}
\mathrm{IPR}&=\left[\sum_i \left( \frac{\Delta \phi_i^2}{\sum_j \Delta \phi_j^2} \right)^2 \right]^{-1}.
\end{split}
\end{equation}
Here $\Delta \phi_i^2/\sum_j \Delta \phi_j^2$ gives the participation fraction of spin $i$ and if the participating spins contribute equally to the transition, the IPR gives their number. On the other hand, the IPR is always greater or equal to 1 and can only be 1 when there is one spin participating. The distribution of IPR obtained by mART using a single spin perturbation is given in Fig.~\ref{fig:IPRAndEnergyBarriers}b (blue curve). It reveals a maximum at about one and a second peak at about two. When $\mathrm{IPR}\approx 1$, there is essentially one spin rotating in a fixed effective field yielding a transition back to the initial configuration. When $\mathrm{IPR}\gtrsim2$, the system can reach a new end configuration, such that it experiences a transition event. 

The peak at two in the distribution of IPRs indicates an important contribution of spin pairs as collective variables for the transition events. The natural question is why are pairs particularly involved in transitions triggered by a single spin. To answer this, we note that the structural arrangement obtained from a Poisson point process in 2D results in the occurrence of nearest neighbour pairs (NNPs), i.e. pairs of spins that are mutually nearest neighbours~\cite{Pickard_1982}. In particular, a spin selected at random has a 62\% chance to belong to an NNP and 59\% in our finite sample (NNP bonds are represented in Fig.~\ref{fig:spinGlass}b). Moreover, NNPs are likely to evolve locally (see appendix~\ref{app:NNP}), i.e. to minimize predominantly their own interaction. Consequently, when one spin of an NNP is perturbed, the second is likely to accommodate it, which activates the whole pair at the transition and gives $\mathrm{IPR}\gtrsim2$. 

We find that 97\% of the transitions obtained by initially perturbing a spin from an NNP have an $\mathrm{IPR}>1.5$. In contrast, only 28\% of the transitions obtained by perturbing a spin that does not belong to an NNP have an $\mathrm{IPR}>1.5$. The difference between these two fractions gives a first evidence for the presence of NNPs as the origin of the large majority of the collective transition events ($\mathrm{IPR}>1.5$). 

The above sampling of the transition states by mART is restricted by the number of spins present in the system (284 different transition states for 800 trial directions), which limits the amount of statistics that can be extracted from the procedure applied on one metastable state. Alternatively, one may choose random directions in phase space, which allows for an unlimited number of attempts~\cite{Mousseau2012}, or employ the 800 directions contained in the 400 harmonic modes of the initial state. The latter has again the drawback to be limited in the number of probes. We set up mART with both these types of perturbation to find 243 transitions from 4468 random perturbations and 349 transitions from the 800 harmonic mode perturbations. Both are expected to have biases that cannot be easily predicted~\cite{Mousseau2000}. In Fig.~\ref{fig:IPRAndEnergyBarriers}b, one can observe that, while the single spin perturbation is biased to sample single spin transitions, the random perturbation tends to find more collective transitions, increasing slightly the tail of the energy barrier distribution (Fig.~\ref{fig:IPRAndEnergyBarriers}a). In absolute numbers however, the harmonic mode perturbation is the most efficient way of sampling both non-collective and collective behaviour. Combining the results of the three methods gives a total of 482 different transitions.  

With this more complete set, we characterise the transition events ($\mathrm{IPR}>1.5$) more precisely. Motivated by the relevance of the NNPs, whose local behaviour arises from their heavy-tailed bond statistics (see appendix~\ref{app:NNP}), we determine the pair which contributes most to the barrier energy of each transition, i.e. the pair $\{k>l\}$ such that $\delta E_{kl}=\max_{i>j}\left(\delta E_{ij}\right)$ for 
\begin{equation}\label{eq:mostContributing}
    \Delta E=\sum_{i>j} \delta E_{ij}\;.
\end{equation}
The spatial distribution of these pairs is identified through their bond in Fig.~\ref{fig:spinGlass}c. We note that many of these bonds do indeed correspond to NNPs (Fig.~\ref{fig:spinGlass}b). In particular, 70\% of the most contributing bonds of all transition events are made of NNPs. When this is not the case, such a bond usually involves one spin of an NNP and its second nearest neighbour.

\begin{figure}[h]
    \includegraphics[width=1.0\linewidth,trim=0.1cm 0.4cm 0.1cm 0.3cm, clip]{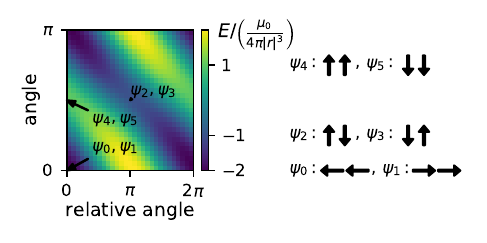}
     \caption{\label{fig:2spinsMEL} (left panel) Magnetic energy landscape (MEL) of two spins with dipolar interaction, plotted as a function of the angle with respect to the displacement vector (angle) and the relative angle between the two spins (relative angle). The MEL is folded according to the symmetries of the spin pair, such that degenerate configurations occur at the same location. (right panel) Corresponding spectrum of states including the Ising-like ground states ($\psi_0$ and $\psi_1$), the two states of the low energy transition channel ($\psi_2$ and $\psi_3$) and the two states of the high energy transition channel ($\psi_4$ and $\psi_5$).}
\end{figure}

\begin{figure}[h]
\includegraphics[width=1.0\linewidth,trim=0.1cm 0.1cm 0.4cm 0.1cm, clip]{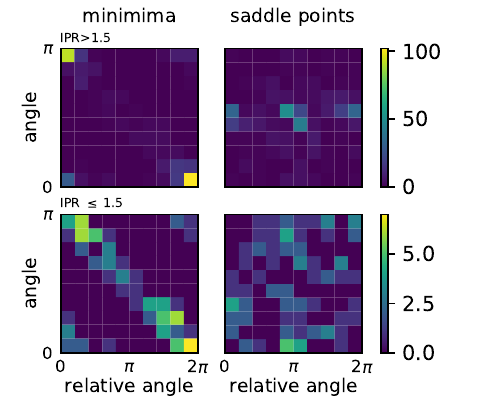}
\caption{\label{fig:isolcatedPairsConfigurations}
2D histograms of the configurations of the most contributing pairs at the initial metastable state and at the transition state for all the transitions. The axes are the same as in Fig.~\ref{fig:2spinsMEL}. The two histograms at the top represent the pair configurations for transition events with $\mathrm{IPR}>1.5$. Similarly, the two bottom histograms are for $\mathrm{IPR}\leq 1.5$.}
\end{figure}

To investigate further the behaviour of these most contributing pairs, we first consider the MEL of an independent spin pair with dipolar interaction in Fig.~\ref{fig:2spinsMEL}: it has a low energy transition channel (states $\psi_2$ and $\psi_3$), and a three-times-higher energy transition channel (states $\psi_4$ and $\psi_5$). For comparison, the distributions of configurations of the most contributing pairs at the metastable state and at the transition state are plotted in Fig.~\ref{fig:isolcatedPairsConfigurations}. The most contributing pairs are mostly populating the two local transitions channels, with a slight imbalance towards the lowest energy transition state. This demonstrates that they indeed evolve locally during the activated transitions of the glass. It is important to add that this result still holds if considering only transitions with $\mathrm{IPR}>2.5$ (not plotted), suggesting that the most contributing pairs still excite independently of the other active spins. The same distributions for transitions with $\mathrm{IPR}\leq1.5$ are also given as benchmarks, showing that the configurations deviate from the predictions for an independent pair.

\begin{figure}[h]
\includegraphics[width=1.0\linewidth,trim=0cm 0.6cm 0.3cm 0.4cm, clip]{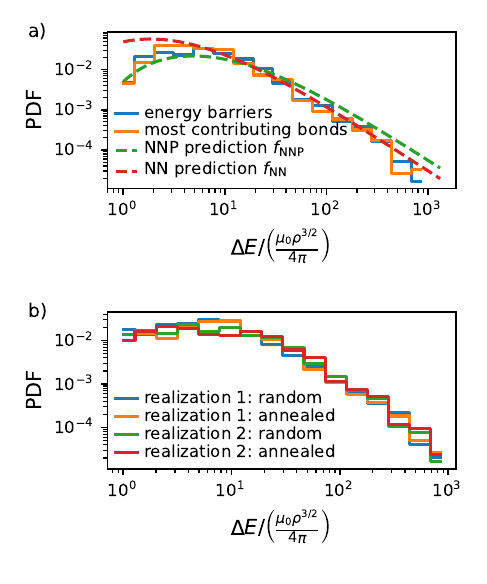}
\caption{\label{fig:EnergyBarriers} a) Log-log plot of the distribution of energy barriers for the total of 387 different transitions with $\mathrm{IPR}>1.5$, as well as the distribution of energy changes for the most contributing bonds (Eqn.~\ref{eq:mostContributing}). In addition, the probability density function for the transition energy related to activating an NNP $f_{\mathrm{NNP}}$ (Eqn.~\ref{eq:E_NNP}), as well as the one related to activating a spin and its nearest neighbour $f_{\mathrm{NNP}}$ (Eqn.~\ref{eq:E_NN}). b) Log-log plot of the distribution of energy barriers for the transitions with $\mathrm{IPR}>1.5$ about a metastable configurations for two different realizations of quenched disorder and obtained either from a random magnetic state or a simulated annealing procedure.}
\end{figure}

We have now understood the implication of NNPs in the most contributing pairs directly (70\% of the most contributing bonds are NNPs) and indirectly (most contributing pairs often evolve locally, likewise NNPs). By laying the distribution of normalised energy change for the most contributing bond over the glass energy barrier distribution in Fig.~\ref{fig:EnergyBarriers}a, we further show that the most contributing pairs dominate the transition energies, downgrading the interaction energy with the environment to a higher order. Using the identification of the most contributing pairs as NNPs, this last result may allow us to develop an analytical understanding of the energy barrier distribution, by developing the interaction statistics of the NNPs. In particular, based on the underlying Poisson point process in 2D, the distance $X$ between the spins of an NNP follows a Weibull distribution~\cite{Pickard_1982}. Translating this for the NNP dipolar interaction energy normalised by the energy scale, i.e. $U=\rho^{-3/2}X^{-3}$, we obtain the following probability density function:
\begin{equation}
f_U(u)=w\left(u; \alpha=2/3, \beta=6\right)\;,
\end{equation}
where
\begin{equation}\label{eq:weibull}
w(u; \alpha, \beta)=\alpha \beta u^{-\alpha-1}\exp(-\beta/u^{\alpha})
\end{equation}
is an inverse Weibull distribution whose shape is controlled by the two parameters $\alpha$ and $\beta$. More specifically, the probability density function for the transition energy of an NNP, assuming an even distribution between the two transition channels (Fig.~\ref{fig:2spinsMEL}), writes
\begin{equation}\label{eq:E_NNP}
f_{\mathrm{NNP}}(v)=0.5w\left(v; 2/3,6\right) +0.5w\left(v;2/3,6\times3^{2/3}\right)\;.
\end{equation}
This analytical prediction for the transition energy is also represented in Fig.~\ref{fig:EnergyBarriers}a. The agreement with the energy barrier distribution is good, confirming both the close relationship between the most contributing pairs and the NNPs as well as the leading nature of their interaction. Limitations of this result can be seen in the lighter tail of the numerical distributions compared to the analytical NNP prediction, entailing possibly the participation of non-NNP bonds (30\% of the most contributing bonds are not NNPs), which should have less extremal statistics. For example, note that the probablitiy densitiy function for the normalised bond energy change in a transition between a spin and its nearest neighbour (which is not necessarily a mutual nearest neighbour) is
\begin{equation}\label{eq:E_NN}
f_{\mathrm{NN}}(v)=0.5w\left(v; 2/3, \pi\right)+0.5w\left(v; 2/3, \pi\times3^{2/3}\right)\;.
\end{equation}
The scales $\beta$ of these inverse Weibull distributions are approximately twice as small as in $f_{\mathrm{NNP}}(v)$, which favours a lighter tail and seems indeed to approximate slightly better the tail of the energy barrier distribution (Fig.~\ref{fig:EnergyBarriers}a).

To complete this analysis, we perform the same procedure for another realization of the quench disorder and find the same activation of the most contributing pairs identified as NNPs, resulting in a very similar energy barrier distribution (Fig.~\ref{fig:EnergyBarriers}b). This is not surprising, as the transitions, being localized in a small region of the sample, already probe different independent realizations of the disorder. More interestingly, we also repeat the analysis for a metastable state of the two quenched disorders obtained after a simulated annealing routine with a linear temperature schedule composed of $5\times10^5$ sweeps starting at temperature 10 times larger than the maximum energy barrier found down to a temperature 100 times smaller than the minimum energy barrier. We find that the energy of the metastable state is not significantly changed and that the conclusion is the same as before (Fig.~\ref{fig:EnergyBarriers}b), indicating that the topology of the MEL is not sensitive to the annealing protocol and remains dominated by the NNPs.

\section{\label{sec:discussion}Discussion and concluding remarks}
mART allows for an efficient navigation through complex energy landscapes to locate transition states. The algorithm proceeds in a controlled and efficient way thanks to a simple failure criterion (section~\ref{sec:failure}) and an adaptive step magnitude (section~\ref{sec:stepMagnitude}). In addition, it is very versatile with a fixed set of parameters ($\gamma=1$ and $\epsilon=0.1$), as shown by the successful applications on different systems. 

Nevertheless, mART cannot be used autonomously since one needs to decide on a set of perturbations to initialize the search in the high dimensional landscape. On the one hand, for systems with an ordered ground state, as studied in the validating examples (section~\ref{sec:benchmarking}), the low energy transitions are affecting a region as large as a the characteristic magnetic lengthscale and it is sufficient to homogeneously perturb on such a scale. On the other hand, for disordered systems lacking symmetries, such as the dipolar spin glass in section~\ref{sec:glass}, there is no discriminating criterion to distinguish a relevant sector of configurations in which to look for transitions. In such a case, following random directions or the harmonic modes are two valid options. For the case of our dipolar glass, we note that it requires more than six times the number of attempts with the random perturbations to recover as many transitions as with the harmonic mode perturbations. This is in contrast to structural work of Mousseau {\em et al.}~\cite{Mousseau2012} which found that the harmonic modes do not contain information about nearby transition states. Quite differently for the dipolar glass, we observe that there are different orthogonal directions contained in the harmonic modes that lead to distinct transitions. Indeed, the relationship between the initial direction and the transition state configuration is made clear by the spatial overlap between the two. This observation might be typical for systems in which the harmonic mode and the transition state are both well localized --- a common feature of disordered systems. Nonetheless, such disordered systems may present many more transitions adjacent to a metastable state than the number of degrees of freedom. It is therefore anyway statistically relevant to additionally sample along random directions. This will also help to reduce possible bias effects (Fig.~\ref{fig:IPRAndEnergyBarriers}) of either or both techniques. Finally, for this specific disordered system, perturbing independently every degree of freedom is found to be less effective than following the harmonic modes: it has a tendency to capture single spin transitions (Fig.~\ref{fig:IPRAndEnergyBarriers}b), which are irrelevant for the evolution of the system. 

In this paper, we not only validate and discuss the implementation of mART, but in sections~\ref{sec:skyrmions} and \ref{sec:glass} we additionally obtain meaningful physical results.
In section~\ref{sec:skyrmions}, a system on a square spin lattice with a ground state hexagonal skyrmion arrangement emerging from a chiral Dzyaloshinskii–Moriya interaction is considered. Due to the topological nature of the skymrions, such a system inherits the thermodynamic features of particle systems at low temperature. For a parametrisation corresponding to large skyrmions compared to the spin lattice parameter, we confirm the existence of a regime of interacting skyrmions whose transport is facilitated by activated processes typical of 2D particle systems. mART can reveal these processes and their energy barriers, which turn out to be an order of magnitude lower than the one of creating or annihilating a skyrmion or equivalently a topological defect in the skyrmion lattice. It also concludes that the skyrmion interstitials should diffuse faster than the skyrmion vacancies reminiscent of observations for particle systems in past numerical and experimental studies~\cite{Libal_2007, Kim_2020}. Nevertheless, some of the common defects in particle systems are not found to be stable in the skyrmion system limiting thereby the number of diffusion paths. This is attributed to the presence of the underlying square spin lattice, which can destabilize the defect configurations due to its incommensurability with the skyrmion hexagonal lattice. This is not surprising, as the presence of a periodic substrate is known to modify the 2D thermodynamic phases depending on its period and commensurability with the hexagonal lattice~\cite{Nelson_1979}. For this specific skyrmion system, the melting was found to occur as a single continuous phase transition~\cite{Nishikawa_2019}. Overall, mART can recover transitions of the interactive skyrmion regime, which paves the way for understanding even more exotic interacting dynamics induced by explicit disorder in the Hamiltonian, such as the one of skyrmions with defects~\cite{Reichhardt_2022} or the one of a skyrmion glass~\cite{Hoshino_2018}.    

For the 2D dipolar interacting spin glass of section~\ref{sec:glass}, the method unveils a strong participation of spin pairs in the transition events. This can be justified by the activation of an extensive number of nearest neighbour pairs (NNP)s, whose interactions are energetically dominating and evolve therefore as independent degrees of freedom to realize the transitions. In particular, the distribution of energy barriers obtained by mART is compatible with the statistics induced by these NNPs (Fig.~\ref{fig:EnergyBarriers}a). This proves that despite the biases of the method in the activation (Fig.~\ref{fig:IPRAndEnergyBarriers}) and in the systematic search~\cite{Mousseau2000, Mousseau2012}, mART is able to extract relevant statistical features. The physical implications of this finding are twofold. First, the absence of relaxation and the persistence of NNP transitions after annealing of the glass (Fig.~\ref{fig:EnergyBarriers}b) can be interpreted by the exponential number of equivalent metastable states with locally relaxed NNPs. Indeed, due to the extensive number of NNPs ($0.31N$) and their remaining Ising symmetry (Fig.~\ref{fig:2spinsMEL}), the configurational entropy of the states with locally relaxed NNPs is finite: $0.31\log(2)$. Additionally, this exponential scaling suggests the existence of an activated dynamics dominated by NNPs at low temperature. Second, as our model corresponds to the dilute limit of the hard-core point process studied in~\cite{Pastor2008}, a comparison of the energy barrier distributions may reveal interesting aspects. On the one side, the inverse Weibull distribution (Eqn.~\ref{eq:weibull}) is heavy-tailed compared to the exponential fit in~\cite{Pastor2008}. This can be interpreted by the presence of extreme NNP statistics that are discarded by the exclusion disc around every spin in the hard-core point process. On the other side, the analysis in~\cite{Pastor2008} does not result in a vanishing distribution at zero energy as we find here. The presence of soft modes can however only be expected in the dense limit, where the packing of the discs forces a lattice arrangement, which can support ordering and the presence of a continuous degeneracy as in~\cite{Belobrov1982, Zimmerman1988}. Nonetheless, in the intermediate packing density, the distribution of energy barriers for the hard-core point process is difficult to predict analytically: even if we were to assume that the nearest neighbour dictates the shape of the distribution, it is worth noting that hard-core point processes have a distribution of nearest neighbour distances that is not common~\cite{Akram_2016} and can deviate significantly from the inverse Weibull distributions of the point process.  


In conclusion, we have presented an adaptive method to identify adjacent saddle points of stable magnetic states and thus activated transitions in magnetic systems. The technique presently developed is available via GitHub \cite{Bocquet_mART_2023}. Knowledge of such transitions, their energy barriers and thermal pre-factors derived from harmonic transition state theory, will eventually allow for the study of activated dynamics over timescales not normally accessible using the Landau-Lifschitz-Gilbert dynamics.

\begin{acknowledgments}
The present work was supported by the Swiss National
Science Foundation under Grant No. 200021-196970.
\end{acknowledgments}

\appendix

\section{Local behaviour of nearest neighbour pairs}\label{app:NNP}
Based on a previous work of Pickard \cite{Pickard_1982}, we develop here the statistics of dipolar interaction for nearest neighbours pairs (NNP)s in a 2D Poisson point process. We assume an NNP set by a distance $X$. We set $Y$ as the distance to the nearest neighbour of the NNP, i.e. the smallest distance among the two second nearest neighbour distances. We can compare $Y$ and $X$ by the ratio $R=Y/X>1$, which is known as the degree of isolation and is explicitly given in~\cite{Pickard_1982}. We can also define a degree of locality for NNPs made of dipolar interacting spins by comparing the interaction magnitudes within the NNP and between the NNP and its nearest neighbour, i.e. $Y^3/X^3=R^3$. This quantity is distributed as 
\begin{multline*}
    P(R^3>a)=\\
    a^{-2/3}\frac{\pi/3 +\sqrt{3}/8}{\pi/4+\frac{1}{2}\sin^{-1}\left(\frac{1}{2a^{1/3}}\right)+\frac{1}{4}\sin\left(2\sin^{-1}\left(\frac{1}{2a^{1/3}}\right)\right)}\;.
\end{multline*}
The degree of locality is heavy-tailed and has no finite moment. Its median corresponds to $R^3=3.4$ and, in $28\%$ of the cases, the interaction within the NNP is more than ten times larger than the interaction of the NNP with the next nearest neighbour ($P(R^3>10)=0.28$). In such conditions, we can predict that NNPs are likely to evolve locally, i.e. predominantly minimize their bond energy rather than any other interaction with the environment.

\end{document}